\input ./style/arxiv-general.cfg
\documentclass[aoas,MSNbibl,nameyear,dvips]{arximspdf}
\makeatletter
   \@ifpackageloaded{graphicx}{}{\usepackage{graphicx}}
\makeatother
\usepackage{multirow,dcolumn}

\doi{10.1214/15-AOAS836}
\volume{9}
\issue{3}
\pubyear{2015}
\firstpage{1533}
\lastpage{1548}
\docsubty{FLA}

\makeatletter
\newcolumntype{d}[1]{D{.}{.}{#1}}

\makeatother

\begin{document}
\begin{frontmatter}

\title{Using somatic mutation data to test tumors for clonal
relatedness\thanksref{T1}}
\runtitle{Testing for clonal relatedness}

\begin{aug}
\author[A]{\fnms{Irina}~\snm{Ostrovnaya}\ead[label=e1]{ostrovni@mskcc.org}},
\author[A]{\fnms{Venkatraman E.}~\snm{Seshan}\ead[label=e2]{seshanv@mskcc.org}}\\
\and
\author[A]{\fnms{Colin B.}~\snm{Begg}\corref{}\ead[label=e3]{beggc@mskcc.org}}
\runauthor{I. Ostrovnaya, V.~E. Seshan and C.~B. Begg}
\thankstext{T1}{Supported by National Cancer Institute Grants CA124504, CA167237, CA163251, CA08748,
Susan G. Komen for the Cure Foundation Grant IIR12221291, and the Metastasis Research
Center of Memorial Sloan Kettering Cancer Center. This work was partially funded by the Alan
and Sandra Gerry Metastasis Research Initiative.}

\affiliation{Memorial Sloan Kettering Cancer Center}
\address[A]{Department of Epidemiology and Biostatistics\\
Memorial Sloan Kettering Cancer Center\\
1275 York Avenue\\
New York, New York 10065\\
USA\\
\printead{e1}\\
\phantom{E-mail:\ }\printead*{e2}\\
\phantom{E-mail:\ }\printead*{e3}}
\end{aug}

%
\received{\smonth{3} \syear{2014}}
\revised{\smonth{5} \syear{2015}}

\begin{abstract}
A major challenge for cancer pathologists is to determine whether a
new tumor in a patient with cancer is a metastasis or an
independent occurrence of the disease. In recent years numerous studies
have evaluated pairs of tumor specimens to examine the similarity of the
somatic characteristics of the tumors and to test for clonal relatedness.
As the landscape of mutation testing has
evolved, a number of statistical methods for determining
clonality have developed, notably for comparing losses of heterozygosity at
candidate markers, and for comparing copy number profiles. Increasingly
tumors are being evaluated for point mutations in panels of
candidate genes using gene sequencing technologies. Comparison of the
mutational profiles of pairs of tumors presents unusual methodological
challenges: mutations at some loci are much more common than others;
knowledge of the marginal mutation probabilities is scanty for
most loci at which mutations might occur; the sample space of potential
mutational profiles is vast. We examine this problem and
propose a test for clonal relatedness of a pair of tumors from a single
patient. Using simulations, its properties are shown to be
promising. The method is illustrated using several examples from the
literature.
\end{abstract}

\end{frontmatter}
\section{Introduction}\label{sec1}

One of the major routine tasks of cancer pathologists is to
determine if a new tumor identified in a patient with cancer is a
metastasis of the original primary tumor or a completely new, independent
occurrence of the disease. Traditionally this diagnosis has been
accomplished by comparing the gross histologic features of the tumor cells,
but in recent years evidence from genetic markers has increasingly come to
inform this decision. At the molecular level the DNA of individual tumors
is characterized by many somatic changes, including mutations in
individual genes and losses or gains of large segments of DNA (copy number
changes). Two tumors that originally evolved from the same ``clone'' of
cancer cells will thus possess some somatic changes that are
identical. These identical changes will be present in both the
primary tumor and the metastasis that is seeded by the primary. In
contrast, any similarities in mutational or copy number profiles of pairs
of independently occurring cancers must occur by chance.
Consequently, comparison of the DNA profiles for
the extent of similarities in the patterns of somatic changes is a powerful
strategy for determining the diagnosis of a new tumor as independent or as
a clone of the original primary.

Clonality testing of this nature has been studied by
numerous investigators over the past two decades. However, this period has
been marked by rapid changes in genetic technology, and so the kinds of
data available have evolved. Early studies typically involved examination
of a few candidate markers for loss of heterozygosity (LOH),
representing copy number changes in the genetic region of the marker locus
[\citet{Imyetal02}, \citet{Sieetal03}, \citet{Dacetal05},
\citet{Geuetal05}, \citet{Orletal09}]. The LOH profiles would then
be compared to determine if the two tumors shared a clonal origin. Our
group developed statistical tests designed for this comparison and applied
these in studies of melanoma and breast cancer [\citet{BegEngHum07},
\citet{OstSesBeg08}]. However, as the
technology evolved, investigators were increasingly drawn
to the use of genome-wide techniques for this purpose [\citet{Boletal08},
\citet{Giretal09}]. We have also examined in detail this
framework and have developed methods for comparing the genome-wide
copy number profiles for the purpose of clonality testing [\citet{Ostetal10},
\citet{Ostetal11}]. The statistical framework for
formulating the comparison of copy number profiles is radically
different from the comparison of profiles of individual markers of
LOH even though the fundamental goal of testing for clonal origin is
exactly the same. The current era is marked by a further significant change
in technology, the introduction of deep genetic sequencing
[\citet{Matetal14}]. This approach identifies individual
somatic mutations within genes such as single nucleotide variants,
deletions, insertions and other extremely localized events. These
mutations are usually identified by comparing the tumor sample
with a matched normal sample to screen out germ-line variants. Addressing
the problem of clonality testing from sequence data is very
distinct from the challenges presented in our earlier work on LOH and copy
number profiles. In the former setting [\citet{BegEngHum07}, \citet{OstSesBeg08}] we dealt with data on a limited
number of markers where the marginal probabilities of allelic losses could
reasonably be considered to be constant, greatly simplifying the
construction of the test. Our work on copy number profiles was
challenged by the problems of determining the locations of the allelic
changes and then formulating a probabilistic strategy for determining if
the locations of the changes could reasonably be considered to be
identical [\citeauthor{Ostetal10}
(\citeyear{Ostetal10,Ostetal11})]. With deep
sequencing data, the major challenges are different (and imprecisely known)
marginal probabilities of mutations at individual loci and the fact that
the sample space of potential mutations is vast and can be defined
only loosely.

Information on point mutations has gradually become more common in
the clinic in recent years as specific driver genes have been identified,
some of which have important therapeutic implications in that
mutations in these genes may be targets for available drugs that have
efficacy against tumors with these mutations. For
example, mutations in the gene
\textit{EGFR} can be targeted by the drug erlotinib, while
vemurafenib is especially effective against tumors with
mutations in the gene \textit{BRAF} [K{\"{o}}hler and Schuler (\citeyear{KohSch13}),
\citet{JanAtk14}]. Consequently, such mutational data
are likely to become increasingly available routinely to pathologists when
diagnosing a new tumor in a patient with an existing tumor. It is
our impression that pathologists will typically conclude that the tumors
are clonal if they match on a single mutation of this nature. One of the
goals we seek to address in this article is to answer the question: is this
conclusion justified? More generally, we develop a framework for
assessing the evidence for clonal relatedness of two tumors when there may
be one or more mutations observed in each tumor, some matching and some
nonmatching.

\section{Motivating example and methods}\label{sec2}

We introduce the problem in the context of an interesting
recent example published by \citet{Kunetal13}. The data, displayed in
Table~\ref{tab1}, came from a patient with two primary colon cancers, denoted T1 and
T3, and tumors in both the right and left lungs. Areas of the left
lung tumor with distinct histological features were examined separately.
The investigators were interested in whether or not the lung tumors could
be metastases of one or other of the colon primaries. Since mutations in
the gene \textit{KRAS} are common in colon
tumors the investigators performed \textit{KRAS}
mutation testing. They discovered distinct \textit{KRAS}
mutations in the left and right lung specimens, suggesting these tumors are
independent, but noticed that the left lung tumor shared a
\textit{KRAS} G12D mutation with one of the colon
primaries, suggesting that the left lung tumor might be a metastasis of the
T3 colon primary. We know from much previous experience that
\textit{KRAS} G12D mutations occur in about 8\% of colon
cancers. Based on this fact, how strong is the evidence for a
clonal link between these two tumors? Clearly a match is evidence in favor
of clonal relatedness, but \textit{KRAS} G12D is a common
mutation and so it is not unlikely that two tumors might share this
mutation simply by chance. A match at a location that is more
uncommon would provide stronger evidence for clonality. To enhance the
evidence the investigators elected to perform additional targeted next
generation sequencing on a much more extensive panel of genes, and
the remaining mutations detected are also displayed in Table~\ref{tab1}.
Here we see that the colon T3 primary has five additional
mutations detected while the left lung tumor has 4 or 6 additional
mutations depending on the histological region examined. However none of
these additional mutations match the new mutations in the colon
primary. Clearly, the presence of new nonmatching mutations
diminishes the evidence favoring a clonal relationship, but how do we
quantify the negative evidence in these nonmatches with the
positive evidence in the \textit{KRAS} G12D
match?

\begin{table}
\tabcolsep=0pt
\caption{Data From \citet{Kunetal13}}
\label{tab1}
\begin{tabular*}{\textwidth}{@{\extracolsep{\fill}}lcccccc@{}}
\hline
&  & \multicolumn{5}{c@{}}{\textbf{Observed mutations}}\\[-6pt]
&  & \multicolumn{5}{c@{}}{\hrulefill}\\
 & & \multicolumn{2}{c}{\textbf{Colon tumors}} & \multicolumn{3}{c@{}}{\textbf{Lung tumors}}\\[-6pt]
 & & \multicolumn{2}{c}{\hrulefill} & \multicolumn{3}{c@{}}{\hrulefill}\\
\textbf{Mutation} & \textbf{Probability}\tabnoteref{tt1}& \textbf{T1} & \textbf{T3} & \textbf{Right} & \textbf{Left/Tubular} & \textbf{Left/Mucinous}\\
\hline
KRAS G12D & 0.081 & & \checkmark & & \checkmark & \checkmark\\
KRAS G12S & 0.019 & & & \checkmark & & \\
XPA G74V & 0.004 & & \checkmark & & & \\
PIK3CA Q546P & 0.004 & & \checkmark & & & \\
FBXW7 R465C & 0.004 & & \checkmark & & & \\
APC $\mathrm{R}283^{*}$ & 0.004 & & \checkmark & & & \\
APC $\mathrm{R}499^{*}$ & 0.004 & & \checkmark & & & \\
APC $\mathrm{Q}1065^{*}$ & 0.004 & \checkmark & & & & \\
TP53 R158H & 0.004 & \checkmark & & & & \\
BRAF G596V & 0.004 & & & \checkmark & & \\
BAI3 V499L & 0.004 & & & \checkmark & & \\
PIC3C2B S314F & 0.004 & & & \checkmark & & \\
ETS1 K200N & 0.004 & & & \checkmark & & \\
IKZF1 M301I & 0.004 & & & & \checkmark & \checkmark\\
PRKDC R364H & 0.004 & & & & \checkmark & \checkmark\\
ZNF521 L1136V & 0.004 & & & & \checkmark & \checkmark\\
ALK $\mathrm{E}405^{*}$ & 0.004 & & & & \checkmark & \checkmark\\
GUCY1A2 V627A & 0.004 & & & & \checkmark & \\
ACVR2A A62G & 0.004 & & & & \checkmark & \\
\hline
\end{tabular*}
\tabnotetext[1]{tt1}{Mutations that were not observed in TCGA were assigned a marginal
probability of $(a+1)^{-1}$, where $a$ is the number of cases observed in
TCGA.}
\end{table}

\subsection{Assumptions and notation}\label{sec2.1}

The key technical features of the problem are as follows. First,
mutations could occur at a very large number of genetic loci, depending on
the size of the sequencing panel used. We denote this number by $n$.
We denote by $m$ the number of loci at which mutations are actually observed
in either tumor in the case under consideration. The marginal mutation
probabilities at each locus, defined as
${p}_i$ for the probability of a mutation at
the $i$th locus, will generally only be known
approximately for the common hotspot mutations that have been frequently
observed in the past, and must be small for mutations that have either
never been observed or not previously observed prior to occurrence in the
case under consideration. Since a match at a rare mutation is much
less likely to occur by chance the observance of such a match provides
greater evidence for a clonal origin for the tumors than a match at a
common locus, so our testing procedure must recognize this. We
have elected to use data from the National Cancer Institute-sponsored
Cancer Genome Atlas (TCGA) to estimate these probabilities in our examples
[\citet{Kanetal13}]. Specifically, we aggregated frequencies observed
in the TCGA database with data from the study in question to
obtain an estimate. New solitary mutations that were not observed in TCGA
were assigned a marginal probability of
$(a+b)^{-1}$ where $a$ is the number of TCGA cases
from the cancer site under investigation and $b$ is the number of cases in
the study. In our testing procedure we assume that these marginal
mutation probabilities are known exactly. Later we investigate the
consequences of inaccuracies in these estimates. An additional key
assumption in the statistical test in Section~\ref{sec2.2} below is
independence of mutations in different markers. This is not true
for genes that are linked by some known genetic pathways [\citet{Sweetal09}] and we also explore the implications of this simplifying assumption
later in Section~\ref{sec3.3}.

Finally, we assume that matching clonal mutations occur in
the original clonal cell, but that at
some point a cell from this clone travels to another site in the
body and seeds the development of the metastasis. After this, the two
tumors can continue to evolve independently through further
mutation and the development of new dominant clones that contain both the
original set of mutations and additional independent mutations. Thus clonal
tumor pairs will possess identical mutations that occurred during the
initial ``clonal'' phase of development and additional sets of
distinct mutations in each tumor, most of which will be
nonmatching but some of which could be identical by chance. To
model this process we use a parameter,
$\xi$,
that characterizes the probability that a mutation will occur in the clonal phase as opposed to the
independent phase. Thus $\xi=0$ for
independent tumor pairs and $\xi>0$
represents the strength of the clonality signal. Clearly this is likely to
vary from case to case. If $\xi$ is large
then clonal tumors will typically have very similar profiles,
while if $\xi$ is small then independently
occurring mutations will predominate.

\subsection{A test for clonal relatedness}\label{sec2.2}

Let $n$ be the total number of distinct mutations (markers)
that potentially could occur. Let $A$ denote the set of markers at which a
matching mutation occurs on both tumors, let $B$ denote the set of markers at
which a mutation occurs on one tumor but not the other, and let $C$
denote the set of markers at which no mutations are observed. Further let $D$
denote the set of all loci, that is,
$D=A\cup B\cup C$, and let $E$ denote the set of markers
that experience mutations, that is,
$E=A\cup B$. Applying the
Neyman--Pearson lemma the most powerful test statistic for distinguishing clonal versus independent tumor
pairs is of the form
\begin{eqnarray*}
S_u&=&\sum_{i\in A}\log \biggl[
\frac{\hat\xi}{1-\hat\xi}p_i^{-1}+1\biggr]-\sum
_{i\in E}\log \biggl[\frac{\hat\xi}{1-\hat\xi}(1-p_i)^{-1}+1
\biggr]\\
&&{}+ \sum_{i\in D}\log \biggl[\frac{\hat\xi}{1-\hat\xi}(1-p_i)^{-1}+1-
\hat\xi\biggr].
\end{eqnarray*}

The last term is summed over all markers. Hence is a constant, and
the test reduces to weighted contributions from sets $A$ and $B$, the markers
at which mutations are actually observed.

For the generalized likelihood ratio test we use the maximum
likelihood estimate of $\xi$ in the test
statistic. However, construction of a null reference distribution presents
a major challenge. This depends on the distribution of the statistic
generated from the full set of $n$ markers. In practice markers with
mutations will represent only a tiny fraction of $n$, which itself
may be an extraordinarily large number. Consequently it is very appealing
to approach the problem by using a conditional test, conditioned on the set
of markers with mutations observed in one or both tumors, that
is, $i\in E$. In this conditional setting the
likelihood ratio can be expressed as:
%
\begin{equation}
S_c=\sum_{i\in A}\log \biggl[
\frac{\hat{\xi}}{1-\hat{\xi}}p_i^{-1}+1\biggr]-\sum
_{i\in E}\log \biggl[\frac{\hat{\xi}}{1-\hat{\xi}}(2-p_i)^{-1}+1
\biggr].
\end{equation}

The last term is a constant since it is summed over all of the
markers in the reference set for the conditional test. Consequently the
conditional likelihood ratio test statistic depends solely on the
weighted contributions of the markers with matched mutations on both
tumors, weighted by $\log[(\hat{\xi}p_1^{-1}/(1-\hat{\xi}))+1]$, the same weights
used for the matches in the unconditional test. The maximum likelihood
estimate of $\xi$ in the
conditional setting can be obtained by maximizing the likelihood
numerically under the constraint that
$0\leq\hat{\xi}\leq 1$, where, by definition, $\hat{\xi}=0$ if no matches are observed and $\hat{\xi}=1$
if all observed mutations are matched
on both tumors.

The crucial practical advantage of the conditional test is
that we can generate relatively easily a null reference distribution since
the sampling depends solely on the markers with mutations observed in one
or other of the patient's tumors. This number is relatively small in our
examples, but it is likely to be manageable, at most in the
hundreds, even if full genome sequencing is used, based on projections from
the TCGA project. To obtain a reference distribution we generate the
distribution of $S_c$ under the assumptions
that matches occur randomly through independent sampling and that at least
one mutation is observed at each marker in the set $E$. Specifically, let
$q_i$ be the probability that there is a
matching mutation at the $i$th marker given that
the marker is mutated in at least one of the two tumors. Then
under the null hypothesis that the tumors are independent
\[
q_i=p_i^2/\bigl(p_i^2+2p_i(1-p_i)
\bigr).
\]

To simulate the null distribution of
$S_c$ we randomly generate the matches over
the set $E$, that is, for the $i$th marker
of the $j$th of $T$ simulations we generate
$x_{ij}$ as a Bernoulli random variable with
probability $q_i$ where
$x_{ij}=1$ if there is a match and 0 otherwise
Then the test statistic for the $j$th simulation is given
by
%
\begin{eqnarray}
S_j=\sum_{i\in E}\biggl[x_{ij}
\biggl\{\log \biggl(\frac{\tilde{\xi}_j}{1-\tilde{\xi}_j}p_i^{-1}+1\biggr)\biggr
\}- \biggl\{\log \biggl[\frac{\tilde{\xi}_j}{1-\tilde{\xi}_j}(2-p_i)^{-1}+1
\biggr]\biggr\}\biggr]\nonumber\\
\eqntext{\mbox{for }j=1,\ldots,T,}
\end{eqnarray}
where $\xi_j$ is the MLE based on data from the $j$th simulation.
The $p$-value is given by $\sum_{j=1}^T I(S_j>S_c)/T$. The critical value of this
test at the one-sided $\alpha$ level, denoted $k_\alpha$ is the
smallest value of $k_\alpha$ such that $\sum_{j=1}^T I(S_j>k_\alpha)/T<\alpha$.

\section{Statistical properties}\label{sec3}

In the following we use simulations to address several questions
about the preceding testing strategy. First, since the actual
mutational profiles of tumors arise through a random process represented by
unconditional sampling, is the conditional test valid? Further, in
discarding information about the markers that were tested but
exhibited no mutations in either tumor are we materially reducing
the efficiency of the test? Second, given that in practice we must use
estimates of the marginal mutation probabilities of the mutations that are
observed in any tumor pair under consideration,
how sensitive is the test to inaccuracies in these marginal
probabilities? Third, to what extent are the properties of the test
affected by correlations among mutations?

\subsection{Validity and efficiency of
conditioning on observed mutations}\label{sec3.1}

Because of the computational barriers to use the
unconditional test when there are large numbers of markers that could
harbor a mutation, allied to the fact that the unconditional test
represents the gold standard against which to compare the conditional test,
we have constructed simulations in configurations where the number
of mutations observed reflect settings that we believe will be realistic,
but where the total number of markers is chosen to be small enough to
facilitate the comprehensive computation required in the
unconditional setting. Thus we construct simulations in which
${n}=10\mbox{,}000$. We calculate the size and power of the test for
various combinations of the clonality
signal, $\xi$, the number of markers with
mutations observed in either or both tumors $(m)$ and their
associated marginal probabilities
$\{p_i\}$.

\begin{table}
\tabcolsep=0pt
\caption{Validity and efficiency of conditional test}
\label{tab2}
\begin{tabular*}{\textwidth}{@{\extracolsep{\fill}}lccccc@{}}
\hline
\multirow{3}{45pt}[-2pt]{\textbf{Mean \# mutations per tumor}} & \multirow{2}{45pt}[-12pt]{\centering\textbf{Clonality signal}} &
\multirow{3}{45pt}[-2pt]{\centering\textbf{Mean \# matching mutations}} & \multicolumn{3}{c@{}}{\centering\textbf{Frequency of $\bolds{p<}\mathbf{0.05}$}\tabnoteref{tt2}}\\[-6pt]
&&& \multicolumn{3}{c@{}}{\hrulefill}\\[3pt]
 & & & \multirow{2}{70pt}[4pt]{\centering\textbf{Unconditional test (calibrated)}} & \multirow{2}{*}{\textbf{Conditional test}} &
 \multicolumn{1}{c@{}}{\multirow{2}{65pt}[4pt]{\centering\textbf{Conditional test (calibrated)}}}\\[3pt]
\hline
\phantom{0}5 & 0.0\phantom{0} & 0.10 & 0.05 & 0.01 & 0.05\\
\phantom{0}5 & 0.1\phantom{0} & 0.56 & 0.41 & 0.36 & 0.40\\
\phantom{0}5 & 0.25 & 1.32 & 0.72 & 0.65 & 0.70\\
10 & 0.0\phantom{0} & 0.21 & 0.05 & 0.02 & 0.05\\
10 & 0.1\phantom{0} & 1.17 & 0.60 & 0.57 & 0.59\\
10 & 0.25 & 2.60 & 0.92 & 0.89 & 0.90\\
20 & 0.0\phantom{0} & 0.45 & 0.04 & 0.03 & 0.05\\
20 & 0.1\phantom{0} & 2.41 & 0.83 & 0.81 & 0.82\\
20 & 0.25 & 5.28 & 0.99 & 0.99 & 0.99\\
\hline
\end{tabular*}
\tabnotetext[1]{tt2}{Each row of the table involved a simulation with 10,000 markers in which
the reference distribution for the test involved sampling from the null
distribution 5000 times, and in which the test was repeated 1000 times to
estimate the size (when the clonality signal is 0) or power (when the
clonality signal is $>$0). The marginal frequencies were constructed in the
following way: For configurations with five mutations per tumor, 10 of the
loci had a marginal probability of 0.10 and the remaining 9990 had a
marginal probability of 0.0004. For configurations with 10 mutations per
tumor, 20 of the loci had a marginal probability of 0.10 and the remaining
9980 had a marginal probability of 0.0008. For configurations with 20
mutations per tumor, 20 of the loci had a marginal probability of 0.10 and
the remaining 9960 had a marginal probability of 0.00016.}
\end{table}

In Table~\ref{tab2} we present the test characteristics in the setting in
which the assumptions are correct, that is, the mutations
are generated independently and the marginal mutation probabilities used in
the test are accurate. The configurations of marginal probabilities reflect
in general terms the nature of somatic mutation, namely,
a few ``common'' markers with marginal probabilities of 0.1, and a
large number of ``rare'' markers. We generated configurations in which the
mean numbers of mutations observed per tumor are five, 10
and 20, with clonality signals of 0 (null), 0.1 and 0.25. Details of the
actual marginal probabilities are provided in the table footnotes.
The results show that the conditional test is valid. This can be seen in
the column ``Conditional Test'' for null
($\xi=0$) settings, that
is, the size of the test is consistently less than the nominal 5\%
level of the test, regardless of the configuration of marginal
probabilities. To compare the power of the conditional and unconditional
tests we ``calibrated'' the results by randomization to ensure that the
test size is always exactly 0.05. The results show that the
conditional test has slightly lower power than the unconditional test, but
that the bulk of the information appears to be captured by the
conditioning.

\subsection{Inaccuracies in marginal probability
estimates}\label{sec3.2}

Our test depends on specification of the marginal mutation
probabilities for each locus at which a mutation is observed. At this stage
of genomic knowledge we do not have accurate information for ths purpose so
our strategy is necessarily ad hoc. We have addressed the likely
consequences of misspecification from two perspectives.
First, we generated simulations in which we added noise to the marginal
probabilities used in the data analyses. This was accomplished by
perturbing the ``true'' marginal
probabilities $\{p_i\}$, that is,
those used to generate the data,
to $\{p_i^{\ast}\}$, using $\log\{p_i^{\ast}/(1-p_i^{\ast})\}=\log\{p_i/(1-p_i)\}+\varepsilon_i$,
where $\varepsilon_i$ is a random $N(0,0.5)$ error term. These
incorrect frequencies were used both for calculating the test statistic and
for generating the reference distribution. These errors correspond
approximately to the statistical uncertainty in mutation probability
estimates obtained for marginal probabilities of ``common''
markers in the range 0.05--0.10 from sample sizes in the
range of 500--1000, roughly the current state of
knowledge based on data from The Cancer Genome Atlas project [\citet{Kanetal13}]. The results are in the column ``Random
Errors'' in Table~\ref{tab3} and these can be contrasted with the results based on
the true probabilities in the preceding column. The results demonstrate a
very modest anti-conservative trend. A possibly greater concern is the fact
that for the vast majority of potential mutational locations in
the genome no previous mutation has been observed. Consequently each time a
first occurrence is observed we have elected in practice to use a marginal
estimator, $N^{-1}$, where $N$ is the total number of
patients examined to date, including those from publicly
available databases like TCGA. It is highly probable that this will
typically be an overestimate because of the large number of potential
mutations in the genome and the fact that most of the mutations
observed to date have only been seen in a single patient. To address the
impact of this, we constructed simulations in which the marginal
probabilities of all of the ``rare'' mutations used in the test were
overestimated by an order of magnitude compared to the
probabilities used in generating the data. These are displayed in the
column headed ``Rare Mutation Overestimation'' in Table~\ref{tab3}. This phenomenon
makes the test more conservative since overestimation of the
marginal probability reduces the strength of evidence favoring
clonality, but clearly does not threaten test validity. However,
substantial power is still apparent for the kinds of configurations
examined.

\begin{table}
\tabcolsep=0pt
\caption{Sensitivity of the test to inaccuracies in the marginal
mutation probabilities}
\label{tab3}
\begin{tabular*}{\textwidth}{@{\extracolsep{\fill}}lccccc@{}}
\hline
&&&\multicolumn{3}{c@{}}{\textbf{Frequency of $\bolds{<0.05}$}}\\[-6pt]
&&&\multicolumn{3}{c@{}}{\hrulefill}\\
 & & & \multicolumn{3}{c@{}}{\textbf{Test calculated using}}\\[-6pt]
&&&\multicolumn{3}{c@{}}{\hrulefill}\\
\multirow{3}{45pt}[8pt]{\textbf{Mean \# mutations per tumor}\tabnoteref{tt31}} &
\multirow{3}{45pt}[2pt]{\centering\textbf{Clonality signal}}&
\multirow{3}{45pt}[8pt]{\centering\textbf{Mean \# matching mutations}}&
\multirow{2}{55pt}{\centering\textbf{True probabilities}\tabnoteref{tt32}} &
\multirow{2}{45pt}{\centering\textbf{Random errors}\tabnoteref{tt33}} &
\multirow{2}{65pt}{\centering\textbf{Rare mutation overestimation}\tabnoteref{tt34}}\\[10pt]
\hline
\phantom{0}5 & 0.0\phantom{0} & 0.10 & 0.01 & 0.02 & 0.01\\
\phantom{0}5 & 0.1\phantom{0} & 0.56 & 0.36 & 0.37 & 0.33\\
\phantom{0}5 & 0.25 & 1.32 & 0.65 & 0.67 & 0.65\\
10 & 0.0\phantom{0} & 0.21 & 0.02 & 0.04 & 0.00\\
10 & 0.1\phantom{0} & 1.17 & 0.57 & 0.58 & 0.37\\
10 & 0.25 & 2.60 & 0.89 & 0.90 & 0.80\\
20 & 0.0\phantom{0} & 0.45 & 0.03 & 0.06 & 0.00\\
20 & 0.1\phantom{0} & 2.41 & 0.81 & 0.82 & 0.54\\
20 & 0.25 & 5.28 & 0.99 & 0.99 & 0.94\\
\hline
\end{tabular*}
\tabnotetext[1]{tt31}{In all configurations the data are generated using the same set-ups as
described in the footnotes to Table~\ref{tab1} with regard to the marginal
probabilities of the mutations and the clonality signal.}
\tabnotetext[2]{tt32}{Here the test is computed by using the same marginal probabilities as
were used in the data generation.}
\tabnotetext[3]{tt33}{For the purposes of calculating the test statistic and its reference
distribution, the marginal probabilities of the markers
$\{p_i\}$ were perturbed with random errors
to $\{p^*_i\}$, using $\log\{p^*_i/(1-p^*_i)\}=\log\{p_i/(1-p_i)\}+\varepsilon_i$ where
$\varepsilon_i$ is a random $N(0,0.5)$ error term.}
\tabnotetext[4]{tt34}{For the purposes of calculating the test statistic and its reference
distribution, the marginal probabilities of the common markers are assumed
to be correct but the probabilities of the rare markers are overestimated
by a factor of 10.}
\end{table}

\subsection{Impact of correlations between
markers}\label{sec3.3}

We address the influence of negative and positive correlation
between markers separately. It is well known that genes that operate in
carcinogenic pathways will often not be co-mutated in tumors. That is, a
mutation in one gene in the pathway will be sufficient to lead to
the tumorigenic effects needed. An example is the mutual exclusivity of
\textit{BRAF} and \textit{{NRAS}}
mutations in melanomas. As a result, we know that strong negative
correlations between mutations in different genes can occur. To model this
kind of phenomenon we generate data where subsets of the markers
are classified into groups of ``pathways'' such that co-incident mutations
within a pathway are not possible, that is, there is
exclusivity between mutations in each pathway. To accomplish this we
simply generated a single outcome from a multinomial comprising
all the markers in the pathway with an additional cell of the multinomial
representing no mutation. Further details are in footnote 3 of Table~\ref{tab4} with
results presented in the column ``Negative
Correlations.'' These results show that
this phenomenon has little impact on either the size or the power of the
test.

\begin{table}
\tabcolsep=0pt
\caption{Sensitivity of the test to correlations in the markers}
\label{tab4}
\begin{tabular*}{\textwidth}{@{\extracolsep{4in minus 4in}}lcccccc@{\hspace*{8pt}}}
\hline
& & & \multicolumn{4}{c@{}}{\textbf{Frequency of $\bolds{<0.05}$}\tabnoteref{tt41}}\\[-6pt]
& & & \multicolumn{4}{c@{}}{\hrulefill}\\
\multirow{3}{45pt}{\textbf{Mean \# mutations per tumor}$^{\mathbf{1}}$}
&  \multirow{2}{40pt}[-12pt]{\centering\textbf{Clonality signal}}
&  \multirow{3}{45pt}{\centering\textbf{Mean \# matching mutations}}
& \multirow{2}{60pt}[-14pt]{\centering\textbf{Uncorrelated}\tabnoteref{tt42}} &
\multirow{2}{55pt}[-8pt]{\centering\textbf{Negative correlations}\tabnoteref{tt43}} &
\multicolumn{2}{c@{}}{\multirow{2}{54pt}[2pt]{\centering\textbf{Positive correlations}\tabnoteref{tt44}}}\\
&&&&&\multicolumn{2}{c@{}}{\hrulefill}\\
 & & & & & \multicolumn{1}{c}{\textbf{0.3}} & \multicolumn{1}{c@{}}{\textbf{0.9}}\\
\hline
\phantom{0}5 & 0.0\phantom{0} & 0.10 & 0.01 & 0.04 & 0.02 & 0.02\\
\phantom{0}5 & 0.1\phantom{0} & 0.56 & 0.34 & 0.41 & 0.31 & 0.24\\
\phantom{0}5 & 0.25 & 1.32 & 0.64 & 0.66 & 0.62 & 0.46\\
10 & 0.0\phantom{0} & 0.21 & 0.02 & 0.02 & 0.02 & 0.03\\
10 & 0.1\phantom{0} & 1.17 & 0.57 & 0.55 & 0.56 & 0.40\\
10 & 0.25 & 2.60 & 0.87 & 0.87 & 0.84 & 0.67\\
20 & 0.0\phantom{0} & 0.45 & 0.04 & 0.02 & 0.04 & 0.05\\
20 & 0.1\phantom{0} & 2.41 & 0.81 & 0.83 & 0.75 & 0.55\\
20 & 0.25 & 5.28 & 0.98 & 0.99 & 0.98 & 0.87\\
\hline
\end{tabular*}
\tabnotetext[1]{tt41}{In all configurations, the 10,000 markers are generated using the same
marginal probability set-ups as described in the footnotes to Table~\ref{tab2} with
regard to the marginal probabilities of the mutations and the clonality
signal.}
\tabnotetext[2]{tt42}{Here the test is computed by using the same marginal probabilities and
(uncorrelated) data generation as in Tables~\ref{tab2} and \ref{tab3}.}
\tabnotetext[3]{tt43}{In these configurations, negative correlation between ``pathways'' is
generated as follows, designed such that the overall mean numbers of
matching mutations are equivalent to the corresponding uncorrelated
configuration. Common markers are generated in blocks of 10 using a single
draw from a multinomial distribution in each block with 10 mutually
exclusive outcomes and fixed marginal frequency of 0.1 each. One, two or
four such multinomials, respectively, are generated for each tumor under
the three scenarios (mean \# mutations of five, 10 or 20). In addition,
5000 rare markers are generated in 50 blocks of mutually exclusive markers
of size 100 and fixed rare marginal frequencies for each mutation (4/9990,
8/9980 and 16/9960, resp., for the three scenarios). That is, we
generated one draw from each multinomial with 101 potential outcomes where
none of the 100 markers exhibit a mutation when 101th outcome is selected
(probability of the 101st outcome is 9490/9990, 9180/9980 and 8360/9960,
resp., for the three scenarios). For markers that belong to these
multinomial blocks, the clonality status is drawn once for a whole block,
that is, the whole blocks rather than individual mutations are considered
clonal or independent. The remaining 4990, 4980, or 4960 markers in the
three scenarios are independent of each other and of multinomial blocks and
are generated as described above. The test statistic and reference
distribution are calculated assuming all markers are independent.}
\tabnotetext[4]{tt44}{Similar to (3) above, one, two or four blocks of size 10 of common
markers and 50 blocks of size 100 of rare markers are generated with
positive correlation. To accomplish this we generated multivariate normal
variates Y of size 10 or 100 with 0 mean, variance 1 and pairwise
correlations of 0.3 or 0.9. The correlated binary mutation outcomes were
determined by dichotomizing these normal variables at the appropriate
marginal frequencies. Clonality status was drawn on per-block basis as
described in (3) above, and the remaining markers were generated
independently. Note that in this set-up it is possible for greater than one
mutation to be observed within a block (indeed this is increasingly likely
as the correlation increases) while in the mutually exclusive construct in
(3) above at most one mutation is observed in each block.}
\end{table}

From an intuitive standpoint, positive
correlation is a much more problematic phenomenon in principle since it
will lead to a greater chance of pairs of matched loci occurring together,
with the potential to greatly inflate the evidence favoring clonality when
the reference distribution assumes independence of markers. To
address this we generated data where once again the markers were grouped
into ``pathways.'' However, in contrast to the
mutually exclusive set-up above, here sets of markers with pairwise
positive correlations are generated within the pathways with the
same marginal frequencies as for the corresponding ``uncorrelated''
framework. Briefly, markers within groups are generated as multivariate
normal correlated variables which are then dichotomized to produce the
desired marginal frequencies (further details are provided in the
footnotes to Table~\ref{tab4}). When the within-blocks correlation is a moderate 0.3
the impact on the test appears to be a modest reduction in power, but the
power decreases notably when the correlations are high (0.9). The
factors bearing upon this loss of power are complex. The increased tendency
for joint occurrence of clonal pairs of correlated markers has an
anti-conservative influence on the test, but this is offset by the
diminished effective sample size due to the presence of
correlation. Further discussion of this issue is provided in a
supplementary material [\citet{supp}]. We believe
that the scenarios used in Table~\ref{tab4} that suggest the typical overall effect
will be conservative provide a persuasive picture of the likely
impact in practice. To gauge this we examined the empirical correlations
between common genes using the TCGA data for the four major solid tumors:
breast, colon, lung, prostate. That is, we correlated
genes rather than individual mutations: logically the latter
correlations should be lower. We only looked at common genes since for rare
mutations the vast preponderance of mutation pairs have never been observed
to occur together, making it difficult to determine reliably the
likely correlation structure. We cross-tabulated the occurrence of
mutations in all pairs of genes that occur with greater than 10\% frequency
and derived the underlying distribution of pairwise correlations
on the assumption that mutations are binary classifications from
latent correlated normal variates. The 75\% percentile of the pairwise
tetrachoric correlations, computed using the ``polychor'' function from the
R package ``\textit{polychor},'' is
0.19 for prostate, 0.32 for lung, 0.15 for breast and 0.50 for
colorectal, while the maximum values are 0.38 for prostate, 0.59 for lung,
0.35 for breast and 0.86 for colorectal. That is, very high correlations
will only occur occasionally and the preponderance of the gene pairs have
correlations that are sufficiently low to have at worst very
modest impact on the power of the test.

\section{Examples}\label{sec4}

We illustrate the test using examples from the recent literature.
Both studies involved mutational profiling addressing the clonality of
primary metastasis pairs of tumor specimens, as described
earlier in Section~\ref{sec2}. In the example from \citet{Kunetal13} the
investigators performed next generation sequencing on two
colorectal tumors considered clinically to be multifocal and
two nonsmall cell lung cancers, one in the left
and one in the right lung, all identified synchronously in the same patient
(Table~\ref{tab1}). Recall that initially the investigators had only information
regarding a matching mutation between the T3 colon tumor and the
tumor in the left lung at \textit{KRAS} G12D.
Our test based on this single locus has a $p$-value of 0.042. Additional
mutational testing identified five new mutations in the
T3 lesion and either four or six
additional mutations in the mucinous and tubular portions of the
left lung lesion, respectively, though none of these were matched in the
colon versus lung comparisons. By accounting for these additional
nonmatches, our test leads to $p$-values of 0.063
and 0.067 when comparing the colon T3 with the mucinous and
tubular lung lesions, respectively. This shows that
nonmatches contribute small amounts of evidence against clonality
and that this negative evidence will accumulate as more nonmatches
are observed.

We have also analyzed another published example of a
celebrated recent study involving panel sequencing of
nine distinct local foci of prostate cancer, a lymph node
metastasis and a series of distant metastases obtained from autopsy
specimens many years after the primaries were surgically removed [\citet{Hafetal13}]. Mutations were identified in the
\textit{PTEN}, \textit{TP53},
\textit{SPOP} and \textit{ATRX} genes.
The data in Table~\ref{tab5} show that the four metastases are all clearly related
through matches in at least three of these
four genes ($p<0.001$ for tests of any pair of
metastases). Six of the primary specimens have no matches with the
metastases. One primary, denoted P1, matches with the metastases
on the \textit{PTEN}, \textit{TP53},
and \textit{SPOP} mutations ($p<0.001$) and the authors
concluded that this tumor, the least advanced of the nine
primary specimens on a histopathologic basis, contains the lethal clone
that led to the metastases. However, two of the other primaries
have matches with the metastases on the mutation
\textit{SPOP} F133L, and so it is pertinent to address the
statistical evidence that these primaries may be clonally related to each
other and to the metastases. In fact, a
comparison of either P6 or P8 with P1 is significant ($p=0.02$), and a
comparison of either P6 or P8 with any of the metastases is also
significant ($p=0.02$). In short, the authors' interpretation that the low
grade P1 tumor is the sole primary tumor that is related to the
metastases may be an incomplete explanation of the clonal evolution of
these tumors. Interestingly, the local lymph node
possesses none of these mutations and it is entirely possible that it is
linked clonally to some of the other primary tumors through shared
mutations in genes that were not tested. More extensive mutation testing of
additional genes would be needed to fully resolve the clonal development of
these tumors.

\begin{table}
\caption{Data from \citet{Hafetal13}}
\label{tab5}
\begin{tabular*}{\textwidth}{@{\extracolsep{\fill}}lccccc@{}}
\hline
 && \multicolumn{4}{c@{}}{\textbf{Observed mutations}}\\[-6pt]
  && \multicolumn{4}{c@{}}{\hrulefill}\\
\textbf{Site} &\textbf{Tumor} & \textbf{PTEN del.} & \textbf{TP53 R248Q} & \textbf{SPOP F133L} & \textbf{ATRX inversion}\\
\hline
\multicolumn{2}{@{}l}{{Mutation probabilities $\rightarrow$}} & 0.004 & 0.008 & 0.023 & 0.004\\
{Prostate} & {P1} & \checkmark & \checkmark & \checkmark & \\
& {P2} & & & & \\
& {P3} & & & & \\
& {P4} & & & & \\
& {P5} & & & & \\
& {P6} & & & \checkmark & \\
& {P7} & & & & \\
& {P8} & & & \checkmark & \\
& {P9} & & & & \\
{Local Node} & {L1} & & & & \\
{Lung} & {B1} & \checkmark & \checkmark & \checkmark & \\
{Liver} & {M5} & \checkmark & \checkmark & \checkmark & \checkmark\\
{Gastric Node} & {M38} & \checkmark & \checkmark & \checkmark & \checkmark\\
{Lung} & {M40} & \checkmark & \checkmark & \checkmark & \checkmark\\
\hline
\end{tabular*}
\tabnotetext[]{}{Mutations that were not observed in TCGA were assigned a marginal
probability of $(a+1)^{-1}$, where a is the number of cases observed in
TCGA.}
\end{table}

We note that in both these examples several tests were
performed between different tumor pairs from the same patient in order to
explore the possible relationships between individual tumor pairs. We have
not performed any multiple testing adjustments. We note that these tests
are structurally dependent, and indeed the deciphering of the full
set of clonal relationships among multiple tumors in a single case
represents a more complex problem that is beyond the scope of this
article.

\section{Discussion}\label{sec5}

Clonality testing using sequencing is likely to be an
emerging clinical application. It has been known for at least two
decades that more accurate pathological diagnosis of metastases is possible
using mutational testing of tumor samples. However, as yet, routine
mutational testing has not entered the clinic. We are now entering
an era in which routine clinical testing of tumor samples is likely to
become commonplace as oncologists seek to identify actionable
mutations for targeted therapy. In the medium-term the technologies will
involve deep sequencing of panels of genes likely to harbor mutations of
therapeutic potential, such as the one employed in \citet{Wagetal12}. A by-product of such testing will be the availability of
mutational data to test the clonal relatedness of metastases with their
putative primaries. We have provided a testing strategy to perform such
classifications, one that is conceptually and computationally
straightforward to apply, and which appears to enjoy good statistical
properties.

An unusual conceptual challenge in constructing a test in this
context if the fact that the sample space is ill-defined due to
the fact that it is not possible to specify precisely the number
of potential ways in which a DNA mutation can occur. Additionally, for most
realistic gene panels, the number of potential mutations is extremely
large, making it computationally challenging to establish a reference
distribution for any test statistic. We circumvented these
problems by constructing a conditional test, conditional on the actual set
of mutations observed. We showed through simulation that this conditional
test is valid and captures most of the relevant information.

Our examples showed that strong evidence for clonal relatedness is
possible even if matches are observed in only two genes. However, a match
in a single, recurring mutation in major cancer genes such as
\textit{KRAS}, \textit{BRAF},
\textit{PTEN}, etc., may not provide
sufficient evidence to demonstrate clonal relatedness
convincingly. Stronger evidence is possible if the match occurs at a more
rarely occurring genetic locus. If clonality testing were to be performed
routinely in the clinic the gene panel should ideally be
sufficiently large to ensure that several mutations will be
observed in all cases encountered.

A notable limitation to our approach is the fact that we must
assign values to the marginal probabilities at each locus. In our example
we used empirical relative frequencies as estimates for the
marginal probabilities, derived using the
publicly available TCGA data. This raises the question of
how to update these marginal estimates, and in particular how to assign
probabilities for mutations seen for the first time in a new
patient, something that is likely to occur quite frequently. If one uses
smaller marginal probability estimates for nonrecurring mutations
based on the recognition that there are huge numbers of genetic loci at
which mutations can potentially occur then the $p$-values will be
smaller. We advocate an estimation strategy that results in the test being
somewhat conservative, as shown in the simulations. Our simulations of
settings in which we substantially overestimated the marginal
probabilities demonstrated the degree of conservativeness to be
expected if these probabilities are overestimated by an order of magnitude.
As evidence gradually accrues about the frequency of specific mutational
events in cancers, our uncertainty about the
assignment of marginal probabilities will decrease, notably for
the more commonly occurring mutations. Our test is also based on the
assumption that mutational events occur independently at different loci.
While this clearly is not literally true across the
genome, our investigation of the impact of
departures from independence showed that the kinds of dependencies observed
between mutations in the TCGA project are likely to have modest impact on
the properties of the test.

As a final cautionary note it must be recognized that all
mutations from sequencing panels are called after a complex laboratory and
data normalization process that can be influenced by numerous potential
biases, including contamination of the specimen with normal cells and
various laboratory processing artifacts. Also, tumors are
heterogeneous and some mutations may only be present in a small subset of
tumor cells and thus detectable only if sequencing coverage is sufficiently
high. These artifacts can introduce false positives or false
negatives. In short, reaching a definitive diagnosis of clonal
relatedness simply because one of many observed mutations is determined to
be matched in the two tumors may overstate the true strength of the
evidence, even if the matching locus is a ``rare'' locus. However,
with proper curation of the called mutations
false positive matches due to germ-line effects or other
artifacts are unlikely.


\begin{supplement}[id=suppA]
\stitle{Supplementary appendix}
\slink[doi]{10.1214/15-AOAS836SUPP} 
\sdatatype{.pdf}
\sfilename{aoas836\_supp.pdf}
\sdescription{Explanation of results of simulations with correlated markers. In the on-line
supplementary appendix we provide additional calculations and graphs that provide an explanation
of the power trends in the presence of correlated markers..}
\end{supplement}


\printaddresses
\end{document}